# Analytic approach to the one-dimensional spin-Peierls system in the entire frequency range

Ping Sun,[1] D. Schmeltzer,[1] and A.R. Bishop[2]

[1]Physics Department, The City College of CUNY, Convent Avenue at 138th Street, New York, New York 10031
[2]Theoretical Division and Center for Nonlinear Studies, Los Alamos National Laboratory, Los Alamos, New Mexico 87545

(Received 30 March 2000)

We use the two cutoff renormalization group (RG) method to study the spin-Peierls model in one dimension for the entire phonon frequency range. We integrate out the phonon and solve the effective Fermion system via mean field and RG methods. We make use of the symmetry that the Néel, dimerization, and spin current order parameters form an SU(2) triplet based on resolving the fermion into left and right movers. We present the phase diagram and discuss its implications for the organic charge-transfer salt $(TMTTF)_2PF_6$.

It was first pointed out by Peierls[1] that electron-phonon interaction plays a special role in one spatial dimension (1D). For a 1D free electron system, the electronic ground state energy is lowered by a static lattice distortion determined by the Fermi momentum. The gain in the electronic energy is then balanced by the increase in the lattice elastic energy. When the distortion is commensurate, an electronic gap is opened at the Fermi surface and turns the metallic system into a band insulator. This mechanism is found to be effective in other 1D systems, such as the antiferromagnets (AF) (Ref. 2) and ferroelectric perovskites,[3,4] where strong electron-electron correlations are also involved. A recent development is the discovery of the inorganic spin-Peierls compound $CuGeO_3$ in which the transition is related to optical phonons,[5] a mechanism which is not yet adequately understood due to the difficulty that the phonon frequency and the exchange energy are of the same order. In this paper we are going to present an analytic approach to solve this problem in 1D.

Peierls' picture of the adiabatic phonon requires the phonon frequency to be much lower than the electron Fermi energy. The conclusion that the Peierls transition happens for arbitrarily weak electron-phonon coupling is changed if the phonon fluctuation (i.e., nonadiabatic effect) is taken into account. For weak couplings the electronic band gap is not strong enough to overcome the phonon fluctuations. It has been confirmed numerically that the transition always happens at a finite coupling for nonzero phonon frequencies.[6] One of our major goals is to determine the phase diagram at all finite phonon frequencies.[7]

A spin chain can be mapped to a 1D spinless Fermion system via the Jordan-Wigner transform. There are three possible phases: the spin fluid (SF) phase, the Néel ordered spin phase, and the lattice dimerization (DM) phase. These phases are determined by the competition among the kinetic energy (the XY part), the spin-phonon coupling, and the Z-direction exchange. In the low phonon frequency limit, it has been found[2] that if the pure AF Heisenberg chain is in the SF phase, then the spin-Peierls transition still occurs for arbitrarily weak spin-phonon coupling, as for free electrons. In the high frequency limit, a competing next-nearest-neighbor exchange is induced by the phonon.[8] The Peierls transition is found at a finite spin-phonon coupling.[8–12] The phase diagram is determined by the renormalization group (RG) method.[9,8]

Our physical understanding of the Peierls problem is based on the strategy to solve the fast modes while adiabatically freezing the slow ones. The ratio of the characteristic energy of the slow modes to that of the fast modes is then the small parameter in perturbation. The method fails when the mode separation is impossible. This is evident in recent attempts to understand the spin system coupled to phonons with frequencies in the intermediate range, motivated by the experiments on $CuGeO_3$. Numerical methods, including quantum Monte Carlo,[13] density matrix renormalization group,[14–16] and Lanczos diagonalization,[17] have been employed to study the problem. Analytical approaches are attempted from both the low[18] and high[19] frequency sides. However, the problem is intrinsically unsolved.

We suggest an approach which avoids the mode separation and applies to the entire phonon frequency range. We integrate out the phonon degrees of freedom, which is always possible at nonzero phonon frequencies. In the fermionic representation, the resulting system has a nearest-neighbor repulsion and a phonon-induced attraction.[20,21] The phonon propagator [see $D(\tau)$ in Eq. (1) below] is kept fully as long as the frequency is less than a certain cutoff $\Lambda_\omega \sim J$. We apply perturbative RG analysis at weak interactions. If at least one of them is strong, either initially or becoming so as a result of scaling, a mean field method is employed. We are able to determine the entire phase diagram using this approach. We will use our results to understand an interesting recent experiment on the organic conductor $(TMTTF)_2PF_6$,[22] whose phase diagram shows a possible quantum fixed point.[10]

We present our approach by studying the following 1D AF spin system

$$\hat{H} = J \sum_j (\hat{S}_j^x \hat{S}_{j+1}^x + \hat{S}_j^y \hat{S}_{j+1}^y + U \hat{S}_j^z \hat{S}_{j+1}^z)[1 + \gamma(u_{j+1} - u_j)]$$
$$+ \frac{1}{2} \sum_q (|\hat{p}_{2k_F+q}|^2 + \omega_q^2 |u_{2k_F+q}|^2)$$

with $J > 0$, $U \equiv J_z/J \geq 0$, and $\gamma \geq 0$. $u_j$ is the ionic displacement and $u_{2k_F+q}$ its Fourier transform. We consider a dispersionless phonon, $\omega_q = \omega_0$. As a result the phonon can be approximated by an Einstein phonon model.[13,14,16,17] In order not to violate translational symmetry, we must consider $u_j$ as a generalized coordinate and not the physical displacement.

We consider the partition function $Z=\text{Tr}\exp(-\beta\hat{H})$ in the zero temperature limit. The Jordan-Wigner transform (see, e.g., Ref. 9) gives $\hat{S}^x_j\hat{S}^x_{j+1}+\hat{S}^y_j\hat{S}^y_{j+1}\rightarrow-(1/2)(\hat{C}^\dagger_j\hat{C}_{j+1}+\hat{C}^\dagger_{j+1}\hat{C}_j)$ and $\hat{S}^z_j\hat{S}^z_{j+1}\rightarrow(\hat{n}_j-\frac{1}{2})(\hat{n}_{j+1}-\frac{1}{2})$. By using the coherent state functional integral,[23] $Z=\int D[\phi^*,\phi]D[\psi^*,\psi]\exp\{-S[\phi^*,\phi;\psi^*,\psi]\}$ where $\phi_j$ ($\psi_j$) is the coherent state eigenvalue for the phonons (fermions). Following Refs. 6,10,11,21, we integrate out the phonon and get $Z=\int D[\psi^*,\psi]\exp\{-S^{\text{eff}}[\psi^*,\psi]\}$ with

$$S^{\text{eff}}[\psi^*,\psi]\simeq\int d\tau\sum_j\left\{\psi^*_j(\tau)\partial_\tau\psi_j(\tau)-\frac{1}{2}[\psi^*_j(\tau)\psi_{j+1}(\tau)\right.$$
$$+\psi^*_{j+1}(\tau)\psi_j(\tau)]+U\left[\psi^*_j(\tau)\psi_j(\tau)-\frac{1}{2}\right]$$
$$\left.\times\left[\psi^*_{j+1}(\tau)\psi_{j+1}(\tau)-\frac{1}{2}\right]\right\}-\frac{\gamma^2}{8\omega_0}\int d\tau\int d\tau'$$
$$\times\sum_j[\psi^*_j(\tau)\psi_{j+1}(\tau)+\psi^*_{j+1}(\tau)\psi_j(\tau)$$
$$-\psi^*_{j-1}(\tau)\psi_j(\tau)-\psi^*_j(\tau)\psi_{j-1}(\tau)]D(\tau-\tau')$$
$$\times[\psi^*_j(\tau')\psi_{j+1}(\tau')+\psi^*_{j+1}(\tau')\psi_j(\tau')$$
$$-\psi^*_{j-1}(\tau')\psi_j(\tau')-\psi^*_j(\tau')\psi_{j-1}(\tau')], \quad (1)$$

where we have set $J=1$ and discarded the contribution from terms with more than four Fermion operators.[24] $D(\tau)=\theta(\tau)e^{-\omega_0\tau}$ is the phonon Green function with $\theta(\tau)$ a step function. The retardation is characterized by $1/\omega_0$. At high frequencies, $\omega_0>\Lambda_\omega$, the Fermions experience no retardation: $D(\tau)\rightarrow(1/\omega_0)\delta(\tau)$. $\Lambda_\omega\sim 1$ is a frequency cutoff.

We define two order parameters for the Néel and DM phases $\hat{C}^\dagger_j\hat{C}_j-\frac{1}{2}\equiv(-1)^j\hat{\Delta}^{\text{Néel}}_j$ and $\frac{1}{4}(\hat{C}^\dagger_j\hat{C}_{j+1}+\hat{C}^\dagger_{j+1}\hat{C}_j-\hat{C}^\dagger_{j-1}\hat{C}_j-\hat{C}^\dagger_j\hat{C}_{j-1})\equiv(-1)^j\hat{\Delta}^{\text{DM}}_j$. The interactions now become $-U\int d\tau\sum_{j,j'}\Delta^{\text{Néel}}_j(\tau)F(j-j')\Delta^{\text{Néel}}_{j'}(\tau)$ and $-U'\int d\tau\int d\tau'\sum_j\Delta^{\text{DM}}_j(\tau)D(\tau-\tau')\Delta^{\text{DM}}_j(\tau')$, with $F(j-j')=\delta_{j+1,j'}$ and $U'\equiv 2\gamma^2/\omega_0$. When the interaction "$U$" is dominant, the system favors a ground state with Néel order. The ideal case is the state with one particle on every other site, $|\text{Néel}\rangle=\otimes_{j=\text{even (odd)}}|n_j=1,n_{j+1}=0\rangle$ for which $\langle\Delta^{\text{Néel}}_j\rangle_{\text{Néel}}=\frac{1}{2}$ $(-\frac{1}{2})$ and $\langle\Delta^{\text{DM}}_j\rangle_{\text{Néel}}=0$. If the phonon mediated interaction $U'$ dominates, the ground state favors DM order. The extreme case is $|DM\rangle=\otimes_{j=\text{even (odd)}}(1/\sqrt{2})(|n_j=1,n_{j+1}=0\rangle-|n_j=0,n_{j+1}=1\rangle)$ for which $\langle\Delta^{\text{DM}}_j\rangle_{\text{DM}}=\frac{1}{2}$ $(-\frac{1}{2})$ and $\langle\Delta^{\text{Néel}}_j\rangle_{\text{DM}}=0$. To reveal the relation between the order parameters, we separate the Fermions into left and right movers: $\hat{C}_j\rightarrow\sqrt{a}[(-i)^j\hat{\psi}_L(x)+(i)^j\hat{\psi}_R(x)]$ ($x=ja$ with "$a$" the lattice spacing). By keeping only the smooth components, $\hat{\Delta}^{\text{Néel}}(x)=\hat{\psi}^\dagger(x)\sigma_x\hat{\psi}(x)$ and $\hat{\Delta}^{\text{DM}}(x)=\hat{\psi}^\dagger(x)\sigma_y\hat{\psi}(x)$. $\sigma_i,i=x,y$ are the Pauli matrices and $\hat{\psi}^\dagger(x)=[\hat{\psi}^\dagger_R(x),\hat{\psi}^\dagger_L(x)]$. In addition, we can identify the $z$ component with the spin current $:\hat{\psi}^\dagger(x)\sigma_z\hat{\psi}(x):=:\hat{\psi}^\dagger_R(x)\hat{\psi}_R(x):-:\hat{\psi}^\dagger_L(x)\hat{\psi}_L(x):$ and the singlet with the particle density $:\hat{\psi}^\dagger(x)I_{2\times 2}\hat{\psi}(x):=:\hat{\psi}^\dagger_R(x)\hat{\psi}_R(x):+:\hat{\psi}^\dagger_L(x)\hat{\psi}_L(x):$. In the Néel (DM) phase the SU(2) triplet orders in the $x$ ($y$) direction. Both the Néel and DM order parameters are zero on average in the SF phase. In a homogeneous external magnetic field, the singlet couples to the field and its expectation value measures the magnetization.[25]

We introduce an effective field theory description via bosonization, which gives

$$\psi_{L/R}(x)=(1/\sqrt{2\pi\alpha})\exp[\mp i\sqrt{4\pi}\hat{\phi}_\mp(x)].$$

"$\alpha$" is an ultraviolet cutoff. $\hat{\phi}_-$ ($\hat{\phi}_+$) is the bosonic field corresponding to the left (right) movers. The order parameters become[26] $\hat{\Delta}^{\text{Néel}}(x)=-(2/2\pi\alpha)\sin[\sqrt{4\pi}\hat{\phi}(x)]$ and $\hat{\Delta}^{\text{DM}}(x)=-(2/2\pi\alpha)\cos[\sqrt{4\pi}\hat{\phi}(x)]$, $\hat{\phi}(x)=\hat{\phi}_-(x)+\hat{\phi}_+(x)$. We can take $\phi$ to be an angular coordinate in the $x$-$y$ plane, in connection with the SU(2) symmetry discussed before. The Néel phase corresponds to $\langle\sqrt{4\pi}\hat{\phi}(x)\rangle=\pm\pi/2$ and the DM phase $\langle\sqrt{4\pi}\hat{\phi}(x)\rangle=0,\pi$. The competition between the two orders is reflected through the constraint $\hat{\Delta}^{\text{Néel}}(x)^2+\hat{\Delta}^{\text{DM}}(x)^2=1/(\pi\alpha)^2$. There are two stable fixed points corresponding to the DM and Néel phases. In between there is a line of fixed points where the $U(1)$ symmetry in the $x$-$y$ plane is restored. The action Eq. (1) now becomes

$$S^{\text{eff}}=\int d\tau\int dx\frac{1}{2K}\left[\frac{1}{v_F}(\partial_\tau\phi)^2+v_F(\partial_x\phi)^2\right]$$
$$-g_1\int d\tau\int dx\int dx'\sin[\sqrt{4\pi}\phi(x,\tau)]$$
$$\times F(x-x')\sin[\sqrt{4\pi}\phi(x',\tau)]-g_2\int d\tau\int d\tau'\int dx$$
$$\times\cos[\sqrt{4\pi}\phi(x,\tau)]D(\tau-\tau')\cos[\sqrt{4\pi}\phi(x,\tau')], \quad (2)$$

where $F(j-j')\rightarrow F(x-x')=\delta(x-x')+O(a)$. We have defined

$$K=\sqrt{(\pi-U)/(\pi+3U)},$$
$$v_F=(a/\pi)\sqrt{(\pi-U)(\pi+3U)}, g_1=4Ua/(2\pi\alpha)^2,$$

and

$$g_2=(8\gamma^2 a/\omega_0)/(2\pi\alpha)^2.$$

We first solve this action by a mean field method. We assume there is an average value for the field and write $\phi(x)=\bar{\phi}+\delta\phi(x)$. To minimize the free energy, we need $\sin(\sqrt{16\pi}\bar{\phi})=0$ and $(g_2/\omega_0-g_1)\cos(\sqrt{16\pi}\bar{\phi})>0$. Thus we anticipate a Néel (DM) phase when $g_2/\omega_0-g_1<0$ ($>0$). As we shall see shortly, mean field fails to predict the SF phase near $g_2/\omega_0-g_1\sim 0$ for small $g_1$ and $g_2$. In addition, even when it applies, the first order transition predicted by mean field is turned into a continuous one. All these are due to the strong quantum fluctuations.

We can perform a two cut-off RG calculation, following Refs. 10 and 21, when both the interactions are weak. We consider first the region $\omega_0<\Lambda_\omega$:

$$\frac{dK(b)}{d\ln b}=-4\pi K(b)^3\frac{\hat{g}_2(b)\hat{v}_F(b)}{[\omega(b)+\hat{v}_F(b)]^3},$$

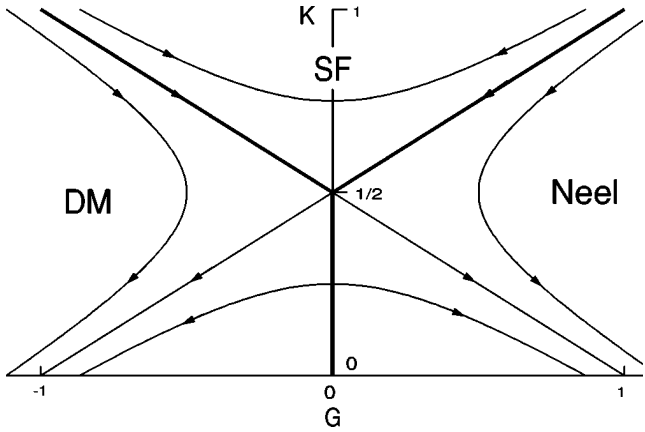

FIG. 1. Phase diagram of the 1D sine-Gordon model near the fixed point. Curves with arrows are a sketch of the RG flows. The heavy lines are the phase boundaries. The SF phase includes the *entire K*-axis plus the region labeled with ''SF.'' All the phase transitions are continuous.

$$\frac{dv_F(b)}{d\ln b} = -4\pi K(b)^2 \frac{\hat{g}_2(b)\hat{v}_F(b)^2}{[\omega(b)+\hat{v}_F(b)]^3},$$

$$\frac{d\hat{g}_1(b)}{d\ln b} = [2-4K(b)]\hat{g}_1(b) + 2K(b)\frac{\hat{g}_2}{\omega(b)+\hat{v}_F(b)},$$

$$\frac{d\hat{g}_2(b)}{d\ln b} = [3-2K(b)]\hat{g}_2(b), \qquad (3)$$

with $\omega_0(b)=\omega_0 b$. We define $\hat{v}_F \equiv v_F \Lambda$, $\hat{g}_1 \equiv g_1/\Lambda$, and $\hat{g}_2 \equiv g_2/\Lambda$ ($\Lambda \sim 1/a$). This set of $\beta$ functions does not show any nontrivial fixed point in the parameter range under consideration. As it is scaled, if either $g_1(b)$ or $g_2(b)$ reaches the order of unity at a length scale $b_0$, while $\omega_0(b_0)<\Lambda_\omega$, the system crosses over to the strong coupling region.[21] If the frequency reaches the cutoff while $g_1(b_0)<1$ and $g_2(b_0)<1$, it crosses over to the high frequency region. The former can be handled by the mean field treatment. For the later, the phonon Green function is effectively a delta function and $S^{\text{eff}}_{\omega_0>\Lambda_\omega} = \int d\tau dx \{[1/(2K)][(1/v_F)(\partial_\tau \phi)^2 + v_F(\partial_x \phi)^2] + G\cos(\sqrt{16\pi}\phi)\}$, $G \equiv \hat{G}\Lambda = (g_1 - g_2/\omega_0)/2$.[10] This is a standard sine-Gordon system and the RG flow is given by[27] $dK(b)/d\ln b = -C\hat{G}(b)^2$ (Ref. 28) and $d\hat{G}(b)/d\ln b = [2-4K(b)]\hat{G}(b)$. The phase diagram is shown in Fig. 1.

This completes the description of our approach to the Peierls problem. For any given system described by a phonon frequency $\omega_0$, spin anisotropy $U=J_z/J$, and spin-phonon coupling $\gamma$, we can determine the physical properties of the system by the above procedure.

We now present the phase diagrams of the 1D spin-Peierls system. We set the nonuniversal constants to be $a = \pi\alpha = 1$, $\Lambda = \Lambda_\omega = 1$. When Eq. (3) applies, we solve the RG flow numerically. In Fig. 2, we show two phase diagrams for given $U$'s. The solid lines are the result of our calculation. We see unphysical cusps on the phase separatrices, which are artifacts of the approximate way to cut off the scaling at $\omega_0 \sim \Lambda_\omega$. A possible (but not unique) way the connect the phase boundaries smoothly is shown by the dashed curves. They are calculated by using a polynomial in the frequency range $0.8 \leq \omega_0 \leq 1.2$ and requiring a smooth connection with the rest. The qualitative feature of Fig. 2(a) has been reached before.[6,13–19] However, since the information from the exact result is used[28] in the calculation, ours is reliable *quantitatively* except in the small frequency region around $\Lambda_\omega$. To see this point, we compare our result with that of Ref. 13, where the critical spin-phonon coupling is calculated accurately to be $\alpha_c = 0.225 \pm 0.015$ at $\omega_0 = 0.25J$ and $U=1$. The parameter $\alpha$ used in Ref. 13 approximately equals $2\gamma$ (it becomes exact when we cosider only the $2k_F$ phonon mode). We have $(2\gamma)_c \simeq 0.210$ at the same $\omega_0$ and $U$. The results coincide. In Fig. 3, the phase diagram for a given frequency $\omega_0 = 0.8$ is shown. Since the frequency here is quite away from $\Lambda_\omega$, RG should give reliable result within the rectangular region (Fig. 3). This diagram shows a similar structure as those sketched in Refs. 8,9 in the high frequency limit. Unfortunately, there lacks reliable numerical result for systems with general values of $U$ with which Figs. 2(b) and 3 can be compared. We shall mention that the special structure of Fig. 2(b) suggests the existence of a re-entrance phenomenon in the frequency direction around $\gamma \sim 0.5$. This may be tested by the other methods, especially numerics. An interesting point of these phase diagrams is that they are de-formations of the sine-Gordon phase diagram. The only exception is that the phase boundary determined by mean field in Fig. 3 is a first order transition. In the following, we argue that the line should represent a coincidence of two continuous transitions while it is in SF phase. Inside the rectangular region in Fig. 3, the phase boundary between Néel and DM phases is a collection of unstable fixed points of SF. In order to connect the line segment of the SF to that of discontinuous transitions, there must be a fixed point. But since the phases on both sides remain the same, this is not true. Besides, at the high frequency limit[9,8] the corresponding line represents a continuous transition. We conclude that all the phase transitions are continuous and lie in the Kosterlitz-Thouless (KT) universality class.

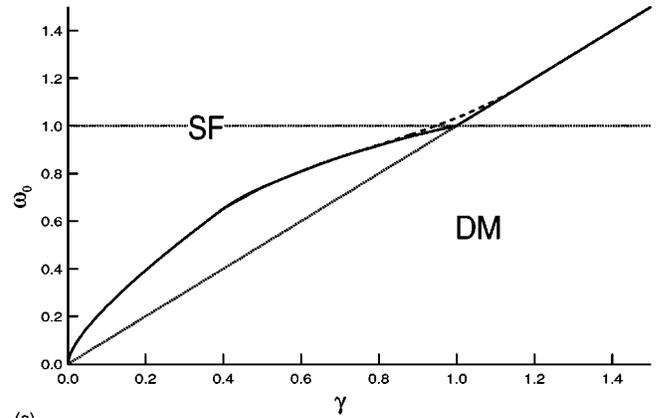

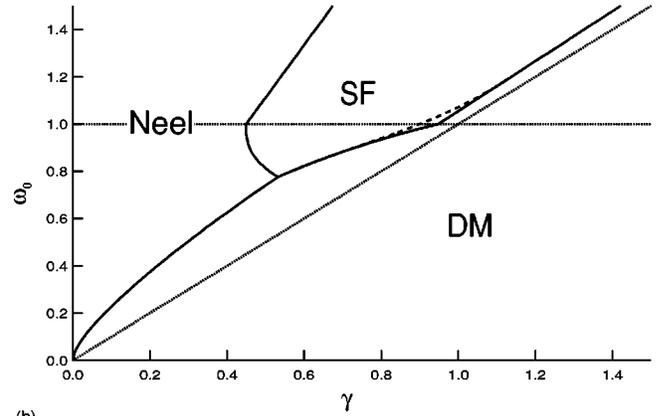

FIG. 2. (a) Phase diagram at $U=1.00$. The phases within the triangular region enclosed by the dotted lines and the $\omega_0$-axis are determined by Eq. (3), while those above the horizontal dotted line by the sine-Gordon result. Mean field is applied to the rest. The solid curve is the calculated phase separatrix. The dashed curve is a fitting as explained in the text. (b) Phase diagram at $U=1.10$. The point where the three curves meet is mapped from the Heisenberg point, $G=0$ and $K=1/2$, in Fig. 1.

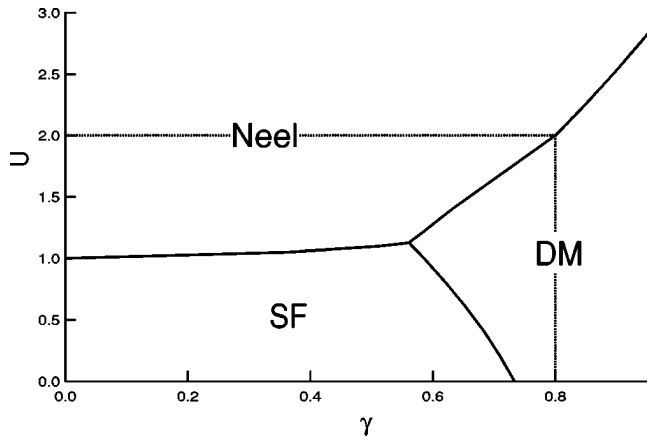

FIG. 3. Phase diagram at $\omega_0=0.80$. Perturbative RG applies within the rectangular region enclosed by the dotted lines and the axes. The mean field result is used outside.

The results presented are useful in qualitatively understanding recent experiments on the 1D organic charge-transfer salt $(TMTTF)_2PF_6$.[22] The experimental phase diagram has been depicted (Fig. 1 in Ref 22) as DM/Néel temperature vs pressure. It shows a similar topology as ours at general values of $U$.[29] The change of pressure results in tuning the spin-phonon coupling and the exchange interactions. Hence it should correspond to a deformed sine-Gordon phase diagram. Although the experiment may not reach it,[29] we expect a quantum critical point where the two phase boundaries merge, as can be seen qualitatively from Fig. 1. We can also conclude that the phase transitions are continuous.

In conclusion, we have presented a unified approach which solves the 1D spin-Peierls system in the entire phonon frequency range. We discussed a hidden SU(2) symmetry to which the order parameters of the Néel, DM, and SF phases belong. We have shown the phase diagrams. We find that all the phase transitions are of the KT type. We suggest quantum critical behavior in $(TMTTF)_2PF_6$. We urge that reliable numerical studies near the frequency cutoff and for general spin anisotropies ($J_x/J$) be carried out.

Work at Los Alamos was supported by the USDOE.